\documentclass{ws-procs975x65}
\usepackage{graphics}
\usepackage{color}
\usepackage{amsfonts}
\usepackage{latexsym}
\usepackage{enumerate}
\usepackage{russ}

\begin{document}

\title{Newman-Penrose-Debye formalism for fields of various spins\\ in pp-wave backgrounds}
\author{Aleksandr Kulitskii$^*$}

\address{Faculty of Physics, Lomomosov Moscow State University,\\ Moscow, Russia\\
$^*$E-mail: av.kulitskiy@yandex.ru}

\author{Elena Yu. Melkumova
}
\address{Faculty of Physics, Lomomosov Moscow State University,\\ Moscow, +74992492878, Russia\\
E-mail: EYM@srdlan.npi.msu.su}

\begin{abstract}
Using Newman-Penrose formalism in tetrad and spinor forms, we perform separation of variables in the wave equations for massless fields of various spins s=1/2, 1, 3/2, 2 on the background of exact plane-fronted gravitational wave metrics. Then, applying Wald's method of conjugate operators, we derive equations for Debye potentials  and we find the back-projection operators expressing multicomponent fields in terms of these potentials. For  shock wave backgrounds, as a special case of the non-vacuum pp-waves, the exact solutions for Debye potentials are constructed explicitly. The possibilities of generalization to the case of massive fields are discussed, in particular, construction of exact solutions of the Dirac and Proca equations.  These results can be used in various supergravity problems on the pp-wave backgrounds, including holographic applications.
\end{abstract}
\keywords{PP-waves, Newman-Penrose formalism, Debye potentials,  shock waves  }
\bodymatter
\section{\label{sec:intro}Introduction}
Wave equations for fields of different spins in curved space have been extensively studied since the early 1970s. Solutions describing  propagation of fields on curved backgrounds  are in demand in  various contexts including the problem of radiation, supergravity and superstrings theories. Generically, considering fields of higher spins in curved space-time, difficulties arise with splitting the systems of coupled equations into separate ones for certain combinations of the field components. Two efficient methods to do this are known. The first is the application  of the Newman-Penrose (NP) formalism \cite{Newman:1961qr}, in which the intrinsic symmetries of the masseless field equations with respect to the Lorentz group are conveniently incorporated. The second efficient tool is the method of Debye potentials \cite{Debye}, which allows to reconstruct the multicomponent fields in terms of solutions of some scalar equations. Debye potentials are complex functions which incorporate two degree of freedom of the massless field of any spin.
 
Formalism of Debye potentials was suggested in gravity theory in the works by Cohen \cite{Cohen:1974cm}  and Kegeles \cite{Kegeles:1979an} and then effectively used by Chrzanowksi \cite{Chrzanowski:1975wv} in Kerr spacetime of type D, according to the Petrov classification \cite{Petrov}, in combination with the Teukolsky equation \cite{Teukolsky:1973ha}. In the Kerr metric Debye representations were found for electromagnetic field and gravitational perturbations. Later, formalism of Debye potentials was discussed in detail for Rarita-Schwinger spin 3/2 field by Torres del Castilho \cite{TorresDelCastillo:1989uq}. 

In this work we apply the Newman-Penrose-Debye formalism to the case of metrics of type N. More specifically, we consider the case of pp-wave metrics, which have numerous  applications both in astrophysics and in various theoretical aspects, including supergravity and holography.

\section{ PP-waves }
These metrics  were introduced by  Brinkmann \cite{Brinkmann:1925fr} in 1925 and  later  interpreted as representing the plane-fronted gravitational waves (pp-waves). They are exact solutions of Einstein's equations of the  following form  
\begin{equation}
ds^2=dudv-H(u,\zeta,\bar\zeta) dudu-d\zeta d\bar\zeta,
\end{equation}
where $\zeta=x+iy$ is a complex transverse coordinate, $H$ is arbitrary nonlinear real profile function of $\zeta$, $\bar\zeta$, which specifies  the nature of the wave. The determinant \mbox{$g=1/16$}  is coordinate-independent. The scalar curvature turns out to be zero.  For $H=0$ the metric is Minkowski. These metrics describe plane waves with parallel rays belonging to the class of solutions that admits  isotropic nonexpanding congruences without shear and twist, as well as the existence of an isotropic Killing vector. 

A natural choice of the  null tetrad basis for them  is
\begin{align}\label{tetrads}
\mathbf l=\sqrt2\partial_v,
&&
\mathbf n=\sqrt2(\partial_u+H\partial_v),
&&
\mathbf m=\sqrt2\partial_\zeta,
&&
\bar{\mathbf m}=\sqrt2\partial_{\bar\zeta}.
\end{align}
We will also use the NP covariant derivatives along the null tetrad vectors:
\begin{align}\label{Differential operator}
&D=l^\mu\nabla_\mu,&&\Delta=n^\mu\nabla_\mu,&&\delta=m^\mu\nabla_\mu,
&&\bar\delta=\bar m^\mu\nabla_\mu
\end{align}
and satisfying, in the case of a plane wave metrics, the following commutation relations 
\begin{align}\label{CommutationOperatorsPPWave}
&[\Delta,D]=0,&&[\delta,D]=0,&&
[\delta,\Delta]=-\bar\nu D,&&
[\bar\delta,\delta]=0.
\end{align}

The nonzero tetrad projections of the traceless part of the Ricci tensor, the Weyl scalar, and the only nonzero spin coefficient $\nu$ are
\begin{align}
\nonumber
&\Phi_{22}=\frac{1}{2}n^\alpha n^\beta R_{\alpha\beta}=-
H_{,\zeta\bar\zeta},
&&\nu=-\bar m^\mu\delta n_\mu= -H_{,\bar\zeta},\\
\label{quantities}
&\Psi_4=-n^\alpha \bar m^\beta n^\sigma \bar m^\tau C_{\alpha\beta\sigma\tau}=-H_{,\bar\zeta\bar\zeta},&&\Lambda=\frac{1}{24}R=0,
\end{align}
where $C_{\alpha\beta\sigma\tau}$ is the Weyl tensor.
In the vacuum case, the gravitational field equations $\Phi_{22}=0$ reduce to the two-dimensional Laplace equation $$H_{,\zeta\bar\zeta}=0.$$
The d'Alembert operator for the massless scalar field $\square=\nabla_\mu\nabla^\mu$ reads:
\begin{equation}\label{KleinGordonOperator}
\square\equiv\frac{1}{\sqrt{|g|}}\partial_\mu\left(\sqrt{|g|}g^{\mu\nu}\partial_\nu
\right)=2(D\Delta-\delta\bar\delta).
\end{equation} 
\section{ Debye potentials}
Here we briefly recall the Wald's procedure \cite{Wald:1978vm} for constructing a solution of the field equations of higher spins, which we then apply to the metrics of plane waves. Consider some multicomponent field $f$ (tensor or spinor) satisfying the field equation  $\mathcal E(f)=0$, 
where $\mathcal E$ is an appropriate matrix linear differential   operator. This is generically a non-quadratic matrix $n\times m$ taking the $m$-component field into the column of $n$ differential equation.  To solve this system of equations one has to disentangle it, which in general is not possible in the closed form. However, it can be possible to decouple a separate equation (or several equations) for some combination $\varphi$  of the components of the initial field $f$ by the action of    another linear partial differential operator $\mathcal T$, defining a scalar $\varphi=\mathcal T(f)$.  Then one can define a pair of linear operators $\mathcal S$ and $\mathcal O$ such, that
$$\mathcal S \mathcal E(f)= \mathcal O\mathcal T(f)=\mathcal O(\varphi).$$ The problem of finding such operators is facilitated if one knows the source terms in the inhomogeneous equations both for an initial multicomponent field $\mathcal E(f)=J$, and in the decoupled scalar equation $O(\varphi)= \mathcal S(J)$. The number of decoupled equations   depends on the number of principal null direction of the metric.

The final step is the construction of the adjoint operator $S^\dag$ with respect to a suitably defined functional scalar product  such that, in the matrix form,
$$\int\phi^{A_n}  M_{A_nB_m}\psi^{B_m}=\int\psi^{B_n} M^\dag_{B_n A_m}\phi^{A_m}\;,$$
where the indices $A_n, B_m$ take $n$ and $m$ values respectively and the integration measure is supressed.   Note that complex conjugation is not used. 

Now it can be verified that the solution of the homogeneous field equation \mbox{$\mathcal E(f)=0$} will be guaranteed if 
\begin{align}\label{DetermineField}
f=\mathcal S^\dag \psi,&&
\mathcal O^\dag\psi=0.
\end{align}
The last equation is the Debye potential equation, the relevant operator is therefore the adjoint
to one used in the equation  for a decoupled scalar. It there are several such decoupled equations (e.g. two in type D metrics), one will have two different representations for the initial multicomponent field which are usually related by some tetrad transformation. In type N case the Debye potential is unique.
 
The complex Debye representation for  for real-valued massless fields reflects the existence of two independent polarizations, which can be  obtained as the real and imaginary parts of the  complex $f$ in the Debye form.

In what follows we apply this procedure to the pp-wave background with the only one nonzero spin coefficient $\nu$.
Clearly, the adjoint to the product of two operators will be the product of the adjoints to each of them in reverse order. So to construct an adjoint of some polynomial product it will suffice to know the adjoint to basic operators.
The adjoint   NP differential operators will read
\begin{equation}\label{adjoints}
D^\dag=-D,\qquad\qquad\Delta^\dag=-\Delta,\qquad\qquad
\delta^\dag=-\delta,\qquad\qquad\bar\delta^\dag=-\bar\delta.
\end{equation}

\section{Maxwell field}
Maxwell's equations in the Newman-Penrose formalism are written in terms of the projections of the electromagnetic field tensor onto the bivector constructed of null tetrad as follows:
\begin{align}\label{MaxwellScalars}
\Phi_0=F_{lm},&&
2\Phi_1=(F_{ln}+F_{m\bar m}),&&
\Phi_2=F_{\bar m n}.
\end{align}
A sourceless decoupled  equation  for  the  scalar  $\Phi_0$  in  the  case  of  vacuum  pp-waves  was obtained in the Refs. \cite{Torres del Castillo:1990} and \cite{Duztas:2015kna}. According to above procedure, we have to construct the source term to this equation. For this, one should act with an appropriate NP differential operators on the first pair of the system  of equations and exclude the variable $\Phi_1$ from them using the commutation relations. This gives 
\begin{align}\label{serapatePhi0}
&\square\Phi_0=2\pi J_0,&& J_0=2\left[\delta J_l-D J_m\right].
\end{align}

Since metrics of type N have only one principal null direction (contrary to the type D, where there are two), the NP component $\Phi_0$ will be the only one for which the decoupling can be made. Further, having written down both adjoint operators, one can easily construct a solution for vector $A^\mu$, satisfying the Lorentz gauge condition, in terms of the Debye potential: 
\begin{align}\label{VectorPotential}
A^\mu=\left[\bar m^\mu D-l^\mu\bar\delta\right]\psi,&&\square\psi=0.
\end{align}
It is easy to see that the contraction of the Ricci tensor with the expression for the vector potential gives  zero; therefore, the solution is valid also for the non-vacuum pp-waves.
\section{Weyl field}
For the Weyl equations of spin 1/2, one uses  the  two-component spinor version of the NP formalism. In this case, we will no longer deal with a tetrad, but with a spinor dyad.  The massless spin 1/2 equation has the following form:
\begin{equation}
\nabla^{A}{}_{B'}\chi_A=0.
\end{equation}
Writing this in components and carrying out transformations similar those for the vector field, we obtain the decoupled equation with the source term:
\begin{align}\label{Spinor1}
&\square \chi_0=N_0, &&N_0=2\left[Dj_{1'}-\delta j_{0'}\right],
\end{align}
where the spinor source is denoted as $j_{A'}$. Then performing the conjugation, we construct a solution in terms of the Debye potential, which, in turn, satisfies the d'Alembert equation:
\begin{align}\label{Spin 1/2 from Debye}
&\chi_0=-D\psi,&&\chi_1=-\bar\delta\psi,&&\square\psi=0.
\end{align}

\section{Rarita-Schwinger field}

The spin 3/2 field is described by the Rarita-Schwinger equation for the
spinor-vector $\psi_\mu$, which satisfies the equations
\begin{align}\label{RSCovariant}
\gamma^\mu(\nabla_\mu\psi_\nu-\nabla_\nu\psi_\mu)=0,&&\gamma^\rho\psi_\rho=0.
\end{align}
This system is consistent only in
the case of the vacuum metrics, the non-vacuum case refers to the supergravity context. 

The field of an arbitrary spin $s$ in the  two-component spinor  formalism is described by a totally symmetric spinor of valence 2s satisfying the equation of motion $$
\nabla^{AB'}\phi_{AB...C}=0.
$$ If $s\geqslant 3/2$ the Buchdahl consistency constraint \cite{Buchdahl:1958xv} $$\Psi^{ABC}{}_{(D}\phi_{EF...)ABC}=0$$ must be satisfyed, where $\Psi_{SABC}$  denotes the Weyl spinor. But there exists also an alternative approach, developed in the work \cite{TorresDelCastillo:1989uq}, where  the consistency constraint is satisfied automatically. This method can be  easily adapted to the case of the  pp-wave backgrounds.

In the spinor equivalent of the Eq. 
(\ref{RSCovariant}) 
\begin{align}
    \nabla_{AB'}\psi^A{}_{CD'}-\nabla_{CD'}\psi^A{}_{AB'}=0,
\end{align}
one has to pass from field $\psi^A{}_{CD'}$ to the symmetric rank three spinor arriving at the modified equations of motion
$$\nabla^A{}^{B'}\phi_{ABC}=
\Psi^S{}_{ABC}\psi^A{}_{S}{}^{B'},\quad \phi^A{}_{BC}\equiv\nabla_{(B|R'|}\psi^A{}_{C)}{}^{R'}.$$ We write down the complete system of equations with the sources for the symmetric spinor field  $\phi_{ABC}$ in the component form,   applicable to the type N metrics:
\begin{flalign}
\nonumber
\;\;\bar\delta\phi_{000}-D\phi_{100}=
\delta J_{0'0'0}
-D J_{1'0'0};
&&\Delta\phi_{000}-\delta\phi_{100}
=\delta J_{0'1'0}-DJ_{1'1'0};&&
\end{flalign}
\\[-35pt]
\begin{flalign}
   \nonumber
&\bar\delta\phi_{100}-D\phi_{110}=
\frac{1}{2}\big[\Delta J_{0'0'0}+\delta
J_{0'0'1}-
\bar\delta J_{0'1'0}
-DJ_{1'0'1}
\big];\\[-4pt]
\nonumber
&\Delta\phi_{100}-\delta\phi_{110}-\nu\phi_{000}
=\frac{1}{2}\big[
\Delta J_{1'0'0}+\delta J_{0'1'1}-\bar\delta
J_{1'1'0}-D
J_{1'1'1}
-\bar\nu J_{0'0'0}\big];\\[-4pt]
\nonumber
&\bar\delta\phi_{110}-D\phi_{111}
=
\Psi_4\psi_{000'}+\Delta J_{0'0'1}
-\frac{1}{2}\bar\delta
(J_{1'0'1}+J_{0'1'1})-\nu J_{0'0'0}
;\\[-4pt]
&\Delta\phi_{110}\!-\delta
\phi_{111}\!-2\nu\phi_{100}\!=\!
\Psi_4\psi_{001'}\!-\!\bar\delta J_{1'1'1}
\!+\!\frac{1}{2}\Delta
(J_{0'1'1}\!+\!J_{1'0'1})\!-\bar\nu J_{0'0'1}\!-\nu
J_{1'0'0}.
\end{flalign}
One can exclude the component $\phi_{100}$ from the first pair of equations,  obtaining the decoupled equation for $\phi_{000}$ with the source term:
\begin{align}\label{phi000 separated}
&\square\phi_{000}=K_0,&&K_0=2\big[D\delta(J_{0'1'0}+J_{1'0'0})-D^2J_{1'1'0}-
\delta^2
J_{0'0'0}\big].
\end{align}
Using this, one can further represent a spin-vector in terms of the Debye potential as follows: 
\begin{align}
\psi_\mu=\left(\begin{matrix}
\bar m_\mu D\bar\delta-l_\mu\bar\delta^2\\
l_\mu D\bar \delta-\bar m_\mu D^2
\end{matrix}\right)\psi,&&\square\psi =0.
\end{align}
The resulting expression  satisfies the system (\ref{RSCovariant}). Note that the equation for the Debye potential is the same  for the vector field, the Weyl field and, as we will see in the next section, also for the tensor field. 
\section{Gravitational perturbations}
Starting with the Einstein's equations and expanding the 
metric on the background 
$$
g_{\mu\nu}=g_{\mu\nu}^{(0)}+h_{\mu\nu},
$$
one derives the Lichnerowitz equation for the spin 2 field in curved space-time \cite{Lichnerowitz}:
\begin{align}\label{Lichnerowitz}
\nabla^\alpha\nabla_\alpha\psi_{\mu\nu}+
2R^\sigma{}_\mu{}^\tau{}_\nu{}^{(0)}\psi_{\sigma\tau}-2
R^\sigma{}_{(\nu}{}^{(0)}\psi_{\mu)\sigma}=0,
&&\nabla_\mu\psi^{\mu\nu}=0,
\end{align}
where $$\psi_{\mu\nu}=h_{\mu\nu}-\frac12g_{\mu\nu}h,\quad  h=g^{\mu\nu}h_{\mu\nu},\quad \psi=g^{\mu\nu}\psi_{\mu\nu}.$$  In the case of non-vacuum background, the additional gauge fixing condition  $\psi=0$ should be imposed.

To apply the NP formalism, similar  splitting has to be performed in the null tetrad vectors, spin coefficients, Weyl scalars and Ricci tensor deviators, equipping the first order perturbations with an index one.  The complete set of gravitational perturbation equations in the Newman-Penrose formalism is obtained by linearizing the Bianchi identities. We present here two equations from the resulting system, which are relevant for pp-waves:
\begin{align}\label{PP1}
&\bar\delta\Psi_0^{(1)}\!-\!D\Psi_1^{(1)}\!=4\pi\big[\delta T_{ll}^{(1)}\!-\!D T_{lm}^{(1)}\big],&&
&\Delta\Psi_0^{(1)}\!-\delta\,\Psi_1^{(1)}\!=4\pi\big[\delta T_{lm}^{(1)}\!-\!D T_{mm}^{(1)}\big].
\end{align}
After some manipulations using   the perturbed  Ricci identities in the NP formalism, we obtain the decoupled equation for the perturbation of $\Psi_0$ with the corresponding source term:
\begin{align}
\label{Spin 2-0}
&\square\Psi_0^{(1)}=4\pi T_0^{(1)},&&T_0^{(1)}=2\left[2D\delta T_{lm}^{(1)}-D^2T_{mm}^{(1)}-\delta^2 T_{ll}^{(1)}\right].
\end{align}
Then using the conjugate operator  we can  write down the solution in terms of the Debye potential:
\begin{align}\label{DerivationPsi}
\psi^{\mu\nu}=\left[2l^{(\mu}\bar m^{\nu)}\bar \delta D-\bar m^\mu \bar m^\nu D^2-l^\mu l^\nu \bar \delta^2\right]\psi,&&\square\psi=0.
\end{align}
This expression satisfies the Lichnerowitz equation with the gauge condition (\ref{Lichnerowitz}).

As in the case of the vector field, this construction will be also valid for the non-vacuum pp-waves. This can be   easily verified by the direct substitution. 

\section{Shock wave backgrounds}

An important subclass of pp-waves is generated via boosting the black hole metrics to an infinite-momentum frame. The line element of the resulting metrics (parametrized by the real transverse coordinates $x_i$ $i=1,2$) reads
\begin{align}
    ds^2=\delta(u)f(\mathrm x)du^2+dudv-dx_i^2.
\end{align}
This is an exact solution of the Einstein equations representing the gravitational shock wave. When $u\neq0$, the  space-time is flat, but for $u=0$ it has a delta-like singularity. But the field equations on such a background are still meaningful due to lack of singularities  in the metric determinant. The function $f (\mathrm x)$ describes the gravitational wave profile. In particular, for the case of the boosted Schwarzschild solution (the Aichelburg-Sexl metric  \cite{Aichelburg:1970dh})
$$f(\mathrm x)=-8p_m\ln\rho,\quad \rho=\sqrt{x_1^2+x_2^2},$$ where $p_m$ denotes the energy of shock wave. For the case of a boosted Einstein-Maxwell-dilaton solution \cite{Cai:1998ii} $$f(\mathrm x)=-8p_m\ln\rho+
\frac{(3-4a^2)}{(1-a^2)}\frac{\pi p_e}{\rho},$$ whetre $a$ - the dilaton coupling constant, $p_e$ - the electric charge. For boosted Taub-NUT \cite{Argurio:2008nb} $$f(\mathrm x)=-8p_n\tan^{-1}\frac{x_1}{x_2},$$ where $p_n$ is the NUT charge. Also known in the literature are the boosted Kerr- Newman solution \cite{Lousto:1992th}, the boosted Schwarzschild-anti-de Sitter \cite{Podolsky:1997ri} space  and some other metrics.

Here we present the solution of the massive Klein-Gordon equation 
on the background of the singular shock-wave metrics: \begin{equation}\label{WaveEquation}
(\square+m^2)\phi=2(D\Delta-\delta\bar\delta)\psi+m\psi=
\left(4\partial_u\partial_v
+4\,\kappa\,\delta(u)f(  \mathrm x)\partial_v^2-\partial_{i}^2+m^2\right)\phi=0.
\end{equation}
In spite of presence of the singularity, there exists an exact  solution of this equation containing the Heaviside function discontinuity only:
\begin{align}
\nonumber
    &\phi=\int\exp[i\,\mathcal W]\,d\mathcal G,&&d\mathcal G=\frac{d\bm q}{(2\pi)^2}\,d\bm x,\end{align}
    \\[-27pt]
    \begin{align}
    \label{W}
    \mathcal W=-\frac{k_v}{2}\big[v-\,\kappa\,\theta(u)f(\mathrm  x')\big]-\frac{(\bm k-\bm q)^2+m^2}{2k_v}u+\bm k{\bm x}+\bm q(\bm x'-\bm x),
    \end{align}
where we have introduced the notation $\bm k=(k_{x_1},k_{x_2})$, $\bm q=(q_{x_1},q_{x_2})$, $\bm x=(x_1,{x_2})$ for the two-dimensional transverse vectors. If we put the mass equal to zero, then we obtain the solution of the equation for the Debye potential.

\section{Massive field with spin $\frac{1}{2}$}
Now consider other massive fields, starting from the spin 1/2. In the two-component notation, the Dirac bispinor equation $i\gamma^\mu\nabla_\mu \psi-\mu\,\psi=0$ with \mbox{$\psi=\left(
\xi^A,
\eta_{A'}
\right)$},
splits into two equations  
\begin{align}
\nabla_{AB'}\xi^A-i\mu\,\eta_{B'}=0,&&
\nabla^{AB'}\eta_{B'}-i\mu\,\xi^A=0.
\end{align}
Their solution in the case of shock-wave backgrounds is a generalization  of the previously obtained expression (\ref{Spin 1/2 from Debye}) for the massless field and can be written in the following form
\begin{align}
\nonumber
&\eta_{1'}=2\!\!\int\!\!
\left([k_{x_1}\!-\!q_{x_1}]+i[k_{x_2}\!-\!q_{x_2}]
-\mu
\right)\exp[i\,\mathcal W]\;d\mathcal{G},&&\eta_{0'}=-\!\!\int\!\! k_v\exp[i\,\mathcal W]\;d\mathcal{G},\\[-5pt]
&\xi^0=2\!\!\int\!\!\Big([k_{x_1}\!-\!q_{x_1}]-i[k_{x_2}\!-\!q_{x_2}]+\mu
\Big)\exp[i\,\mathcal W]\;d\mathcal{G},&&
\xi^1=\!\int\!\! k_v\exp[i\,\mathcal W]\;d\mathcal{G},
\end{align}
where $\mathcal W$ defined in (\ref{W}).

\section{Proca equation}
For the massive spin-1 field the 
gauge invariance of Maxwell's field is broken by the mass term. Instead of the gauge fixing condition, we deal with the Lorentz-like dynamical constraint, so we have two equations:
\begin{align}\label{Proka}
\nabla_\mu F^{\mu\nu}+m^2 A^\nu=0,&& \nabla_\mu A^\mu =0.
\end{align}
The massive vector field is no longer transverse, possessing three  physical degrees of freedom. The first two polarizations in the case of shock wave background are realized by   modified expressions  with Debye potentials (the   real and imaginary parts of the solution):
\begin{align}
A^\mu_{(1,2)}\!\!=\!\!\!\int\!\!\mathcal K^\mu_{(1,2)} \!\exp[i\,\mathcal W]\,d\mathcal G,&&\mathcal K^\mu_{(1,2)}\!=\!\Big\{\!0, 2[k_{x_1}\!\!-\!q_{x_1}]\!-\!2i[k_{x_2}\!\!-\!q_{x_2}]
, k_v, -ik_v\!\Big\}.
\end{align}
For the third polarization we  solve the dynamical constraint acting by the covariant derivative on the massive scalar field, obtaining
\begin{align}
\nonumber
A^\mu_{(3)}\!\!=\!\!\!\int\!\!\mathcal K^\mu_{(3)} \!\exp[i\,\mathcal W]\,d\mathcal G+c.c,
\end{align}
\\[-25pt]
\begin{align}
\mathcal K^\mu_{(3)}\!=\!\Big\{\!k_v,\;\frac{(\bm\!-\!\bm q)^2\!-\!m^2}{k_v}\!+\!k_v
\delta(u)f(\mathrm x)
,\;\,k_{x_1}\!\!-\!q_{x_1},\;k_{x_1}\!\!-\!q_{x_1}\!\Big\}.
\end{align}
It can be seen that this expression  satisfies  the  constraint indeed.  But this expression gives us only one additional polarization, because of real multipliers in front of the exponent. Therefore the complete  solution of the Proca equation in AS metric is a sum $A^\mu=A^\mu_{(1,2)}+A^\mu_{(3)}$. To write down the solution for massless electromagnetic field, it is sufficient to set the mass to zero.

\section{ Conclusions}
This work is devoted to some novel  applications of
the Newman-Penrose formalism and the method of  Debye potentials. Previously this technique was successfully used to construct solutions of equations for massless fields of different spins on the background of vacuum black hole solutions of  Petrov type D. Here it was  applied  to solutions of  type N. Unlike the case D, where the metric has two principal null directions and, accordingly,  two decoupled equations for NP projections can be derived, in the metrics of type N   only one decoupled equation exists. Namely, one can decouple the $\chi_0$-equation for the Weyl field, the $\phi_{000}$-equation for the Rarita-Schwinger field, and the equations for perturbations of $\Phi_0$ and $\Psi_0 $  of the vector and tensor fields respectively. It is still enough to construct the Debye representation  for the solutions obtaining the universal equation  for the Debye potential for all spins. We also managed to   generalize our construction   to the case of  massive fields on the background of shock-wave metrics. Our formulas can be used for quantization in shock wave backgrounds and in some holographic applications. 
 
\section*{Acknowledgments}
The work was supported by the Russian Foundation for Basic Research on the project 20-52-18012, and the Scientific and Educational School of Moscow State University "Fundamental and Applied Space Research".


\end{document}